\documentclass[aps,pra,twocolumn,superscriptaddress]{revtex4-1}

\usepackage{graphicx}
\usepackage{float}
\usepackage[version=3]{mhchem}

\begin{document}
\title{Evidence of sympathetic cooling of \ce{Na+} ions by a \ce{Na} MOT in a hybrid trap }
\author{I.~Sivarajah}
\affiliation{Department of Physics, University of Connecticut, Storrs, Connecticut 06269}
\author{D.~S.~Goodman}
\affiliation{Department of Physics, University of Connecticut, Storrs, Connecticut 06269}
\author{J. E.~Wells}
\affiliation{Department of Physics, University of Connecticut, Storrs, Connecticut 06269}
\author{F.~A.~Narducci}
\affiliation{Naval Air Systems Command, EO Sensors Division, Bldg 2187, Suite 3190 Patuxent River, Maryland 20670, USA }
\author{W.~W.~Smith}
\affiliation{Department of Physics, University of Connecticut, Storrs, Connecticut 06269}
\date{\today}
\begin{abstract}
A hybrid ion-neutral trap provides an ideal system to study collisional dynamics between ions and neutral atoms. This system provides a general cooling method that can be applied to  species that do not have optically accessible transitions, and can also potentially cool internal degrees of freedom. The long range polarization potentials ($V\propto-\alpha/r^4$ ) between ions and neutrals result in large scattering cross sections at cold temperatures, making the hybrid trap a favorable system for efficient sympathetic cooling of ions by collisions with neutral atoms. We present experimental evidence of sympathetic cooling of trapped \ce{Na+}  ions, which are closed shell and therefore do not have a laser induced atomic transition from the ground state, by  equal mass cold Na atoms in a magneto-optical trap (MOT).
\end{abstract}
\maketitle
\section{INTRODUCTION}
\label{sec:Introduction}

	 Hybrid traps are ideal experimental systems for studying collisional dynamics between neutral atoms and atomic or molecular ions. Trapped cold ions are potential candidates in studies related to precision measurements \cite{Blatt:92,Diddams:11}, quantum computing \cite{Zoller:95,Everitt:04} and ultracold quantum chemistry  \cite{Krems:08,Donley:02,Grier:2009, Felix:2011,Rellergert:2011,Schmid:2010,ZipkesPRL:2010,ZipkesNature:2010}. In most of these applications the trapped ions are required to be cooled down to low temperatures to extend their storage  times and spectroscopic resolution. Various cooling mechanisms such as laser cooling \cite{Raizen:1992,Birkl:1992}, resistive cooling \cite{Wayne:95, Herfurth:06}, sympathetic cooling by other co-trapped cold ions \cite{Molhave:2000,Blythe:2005,Larson:1986} or  buffer gas cooling \cite{DeVoe:2009,Schwarz:2008,Major:1968} have been regularly implemented.

	As originally proposed by W.W. Smith, our hybrid trap consists of a magneto-optical trap (MOT) concentric with and encompassed by a linear Paul trap (LPT) \cite{Smith:2003,Smith:2005}. The hybrid trap apparatus has recently been used in several experimental measurements of charge-exchange rate constants \cite{Grier:2009,Rellergert:2011,Felix:2011} and  creating a cold molecular ion source \cite{Sullivan:2011}.  Using localized cold or ultracold atomic gases to cool ions in a hybrid trap has been investigated using cold atoms from a MOT (cooling to a steady state of 200 trapped \ce{Rb+} ions) \cite{Ravi:2012} or a Bose-Einstein condensate (BEC) (for a single ion) \cite{Schmid:2010,ZipkesPRL:2010,ZipkesNature:2010}. This paper describes experimental evidence of effective cooling of an initially large population ($\sim 10^3 - 10^4$) of \ce{Na+} ions by elastic scattering and resonant charge-exchange collisions with cold equal-mass Na MOT atoms in a hybrid trap.

Sympathetic cooling is achieved when the translational kinetic energy of one gas is reduced by elastic, inelastic, and charge-exchange collisions with another colder gas \cite{Larson:1986, Goodman:12}. When collisionally cooling one ionic species of mass $m_I$ with another pre-cooled ionic species of mass $m_C$, experiments \cite{Blythe:2005} and simulations \cite{Zhang:2007} have demonstrated that a wide range of mass ratios $m_I/m_C$ can be cooled using this ion-ion sympathetic cooling technique.

In the case of sympathetic cooling by a neutral buffer gas, as first demonstrated by Major and Dehmelt \cite{Major:1968}, the cooling is limited by the ratio between  the ionic ($m_I$)  and the atomic ($m_A$) masses. Only when $m_I/m_A > 1$ can the cooling  overcome the atom-ion rf heating \cite{Schwarz:2006}. More recent theoretical work suggests that the buffer gas mass ratio may go as low as $m_I/m_A >$ 0.65 \cite{DeVoe:2009}.  However, unlike a buffer gas, which is dilute and extends throughout the trapping region, a MOT is a very dense, localized cloud of neutral atoms. As a result, certain approximations made in Ref.~\cite{Major:1968} that lead to $m_I/m_A > 1$ do not apply \cite{Ravi:2012}. Therefore, ion collisions with the MOT result in little to no atom-ion rf heating until the ions secular motion amplitude is smaller than the radius of the MOT. This allows ions to be sympathetically cooled by equal mass neutral atoms \cite{Ravi:2012,Schmid:2010, Goodman:12}.

Sympathetic  cooling is advantageous because it can be applied to atoms and molecules that do not have optically accessible transitions \cite{Raizen:1992,Birkl:1992, Smith:2005} and has been theorized to be able to cool the internal degrees of freedom of molecular ions \cite{Hudson:09, Smith:2005}. Since \ce{Na+} is a closed shell ion, the conclusions of this study should be applicable to other atomic and molecular ions that do not have a laser induced transition. Lastly, sympathetic cooling with a MOT is useful in that it doubles as an efficient source of atomic or molecular ions \cite{Sullivan:2011} and a refrigerant for those equally massive or more massive ions.

Ion-neutral interactions at low energy are dominated by the long range polarization potential $V\propto-\alpha/r^4$, where $\alpha$ is the dipole polarizability of the neutral species. The collision cross sections between ions and atoms are considerably larger than the cross sections between two neutrals at cold temperatures \cite{Cote:2000,Smith:2005, Makarov:2003}. We previously investigated the feasibility of sympathetic cooling of \ce{Na+} and \ce{Ca+} ions by a Na MOT via \textsc{simion} simulations \cite{Goodman:12}. In this simulation paper, we considered both elastic scattering and resonant charge-exchange for \ce{Na+} cooling by a Na MOT and only elastic scattering for the cooling of \ce{Ca+} by a Na MOT. In both cases effective cooling should be achieved. This paper shows evidence of sympathetic cooling
for the \ce{Na+} case.

It was shown in \cite{Ravi:2012} that resonant charge-exchange can play an important role in equal mass ion-neutral cooling within a hybrid trap.  For our system, the charge-exchange cross section is slightly less but comparable to that of \ce{Rb+}-\ce{Rb} \cite{Cote:2000,Ravi:2012}. The dominant cross section is always elastic scattering within the relevant temperature range \cite{Cote:2000}.  The role of charge-exchange collisions may be even more important to our system compared the \ce{Rb}-\ce{Rb+} experiment, since elastic scattering for \ce{Na} is smaller (the elastic cross section scales like $(\mu \alpha^2)^{1/3}$ \cite{Cote:2000}) .  However, as we found in our simulations for \ce{Ca+}, which only included elastic scattering and demonstrated better cooling than the \ce{Na+} case, charge-exchange is not required to sympathetically cool ions with the hybrid trap. 

This paper is organized as follows: In Sec. II, we discuss our experiment, which includes a description of the apparatus as well as trap loss mechanisms. Trap loss mechanisms are important to understand the results of our experiments, which are presented in Sec. III. We conclude in Sec. IV. 

\section{Experiment}
\label{sec:Background}
\subsection{Apparatus}
\label{sec:Apparatus}

	The hybrid trap consists of a \ce{Na} MOT concentric with an ion cloud confined within a linear Paul trap (LPT).

\begin{figure}[t]
   \centering
\scalebox{1.1}{\includegraphics[width=3.1in]{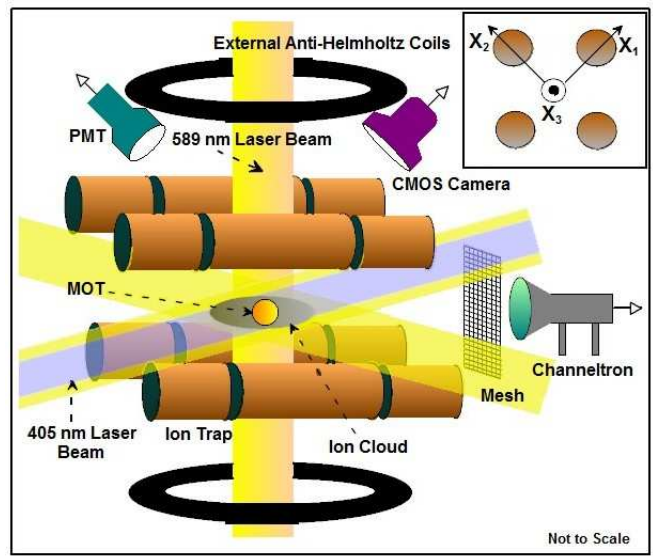} }
   \caption{(Color online).  Diagram of the hybrid trap apparatus. A \ce{Na} MOT (orange) is formed concentric with an ion cloud (grey) inside a segmented linear Paul trap (LPT) with six 589 nm MOT beams (yellow) and a pair of anti-Helmholtz coils (exterior to the chamber). A 405 nm beam (blue) aligned co-linearly with one of the MOT beams is used for REMPI. Fluorescence measurements of the MOT can be made with a photo-multiplier tube (PMT) or a CMOS camera. An electrically biased mesh is placed between the LPT and the Channeltron electron multiplier (CEM) which is used for ion detection. Inset: Axial view of Paul trap with Cartesian coordinate system.}
   \label{fig:Hybrid1}
\end{figure}

	A standard Na type II  MOT \cite{Raab:1987, Tanaka:2007,Prentiss:1988, Goodman:12} is formed as shown in Fig.~\ref{fig:Hybrid1}. In addition to the anti-Helmholtz coils required for the MOT, magnetic shim coils are also placed outside the vacuum chamber, enabling the MOT to be translated for better overlap with the ion cloud.  A \ce{Na} source (Alvatec or SAES) inside the vacuum chamber provides the $\approx \mathrm{1000~K}$ background Na gas from which the MOT is produced. The vacuum chamber was maintained at a constant pressure on the order of $\sim 10^{-9}$ Torr by continuous pumping with an ion pump. Using standard fluorescence measurements taken with a photo-multiplier tube (PMT) and/or a Complementary Metal Oxide Semiconductor (CMOS) camera \cite{Townsend:1995, Foot:92} a peak  MOT density of $\sim 10^{10}~\mathrm{cm^{-3}}$ was inferred for the type II MOT.

	 The release and recapture \cite{Lance:10} technique was implemented to measure the MOT temperature, which was found to be $\mathrm{0.2(1)~mK}$. The model typically used with this measurement assumes ballistic expansion of the MOT cloud, which in a hybrid trap is impeded by the LPT apparatus. Atoms are reflected back into the recapture region by the trap electrodes, systematically lowering the effective temperature measured. Therefore, the measurement was considered a lower limit of the actual MOT temperature, which is likely an order of magnitude higher \cite{Prentiss:1988, Oien:1997}.

	 The \ce{Na+} ions necessary for this experiment are captured by utilizing an LPT. Our LPT consists of four segmented metal rods assembled as shown in Fig.~\ref{fig:Hybrid1} and discussed in further detail in Ref.~\cite{Goodman:12}. The four center segments, termed rf segments, are used for radial confinement of charged particles, while the eight end segments (four on each end) are used for axial confinement. The rf segments are supplied with a voltage of the form $\pm V_\mathrm{rf}\cos{\Omega t}$, with the voltage on the pair of rods along the $x_1$ axis  $180^\circ$ out of phase from the voltage on the pair along the $x_2$ axis. This configuration effectively creates  a rotating quadrupole  saddle potential at the center of the LPT \cite{Raizen:1992, Major:2004,Drewsen:2000,Ryjkov:2005,Paul:90}. Unless otherwise specified, the amplitude was set to $V_\mathrm{rf}=36~\mathrm{V}$ and the driving frequency was set to $\Omega/2\pi=729~\mathrm{kHz}$. Confinement along the axis is provided by a DC potential $V_\mathrm{end}$ = 35 V on the end segments with the center segments at dc ground.

	The total time dependent electrical potential near the center of the LPT due to these applied fields is aproximated by (for $x_1^2 + x_2^2 \ll r_0^2$)
\begin{eqnarray}
\Phi(x_i,t) &\approx&   V_{\mathrm{rf}} ~\mathrm{cos} \left (\Omega t \right ) \frac{x_\mathrm{1}^2-x_\mathrm{2}^2}{r_0^2}\nonumber\\
&&
+\frac{\eta V_{\mathrm{end}}}{z_0^2} \left (x_3^2-\frac{x_\mathrm{1}^2+x_\mathrm{2}^2}{2} \right )
\label{PTpotential}
\end{eqnarray}

 where $x_i$ is the magnitude of the position vector with the coordinates given in Fig.~\ref{fig:Hybrid1} (inset), the distance between two diagonal electrodes is $2r_0$= 19 mm, the length of the rf segment is $2z_0$= 48 mm and $\eta$=0.14 is a unitless efficiency factor dependent on the geometry of this particular trap.

 The motion of a single ion within this oscillating electric field is described by the Mathieu equation \cite{Major:1968}. This motion can be divided into a slow secular motion and a rapid micromotion at the rf driving frequency $\Omega$  \cite{Berkeland:1998}.  

\begin{figure}[t]
   \centering
\scalebox{1.1}{\includegraphics[width=3.3in]{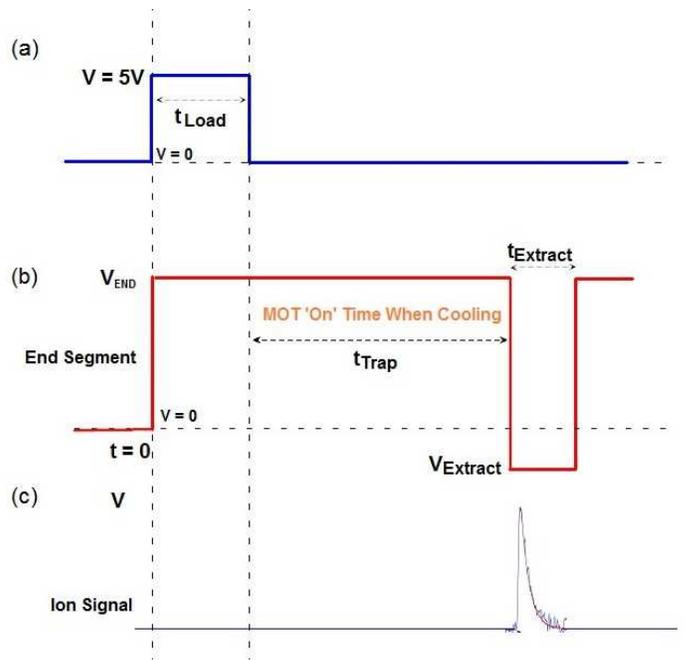} }
   \caption{(Color online). (a) The transistor-transistor logic (TTL) pulse used with the 405 nm laser diode to load (loading time denoted by $t_\mathrm{Load}$) \ce{Na+} ions via the REMPI method. (b) $V_\mathrm{end}$ of the four end segments closest to the CEM is lowered at the time of extraction to generate a dipolar field between the ends of the LPT to direct the ions towards the CEM. (c) A typical ion signal from the CEM and preamplifier.}
   \label{fig:Sequence}
\end{figure}

	  The sequence of loading, trapping and detecting the ions is depicted in Fig.~\ref{fig:Sequence}. Ions were loaded into the LPT via a resonance enhanced multiphoton ionization (REMPI) method \cite{Compton:1980}. A  laser  diode at 405 nm (RGBLase LLC, 100 mW) drives excited  \ce{Na} atoms into the continuum, thereby producing \ce{Na+} ions. The 405 nm laser beam is colinear with one of the MOT beams (Fig.~\ref{fig:Hybrid1}). The initial number of ions loaded within the LPT can be controlled by adjusting either the 405 nm laser intensity or the exposure duration $\mathrm{t_{Load}}$.

	A destructive ion detection method was employed using a Channeltron electron multiplier (CEM) positioned adjacent to the LPT along its trap axis as shown in Fig.~\ref{fig:Hybrid1}.  An electrically biased mesh placed between the LPT and the CEM isolated the CEM from the trapped ions thereby ensuring that the trapping fields were not disturbed by the high operating voltages of the CEM.

	 The trapped ions are extracted by lowering $V_\mathrm{end}$ closest to the CEM from the initial 35 V to -7 V and the opposite $V_\mathrm{End}$ from 35 V to 0.5 V producing a dipolar electric field between the two sets of end segments accelerating and directing the ions into the CEM \cite{Hashimoto:2006,Ravi:2012}. The CEM signal is fed through a preamplifier which outputs the integrated ion signal. The amplitude (peak value) of the ion signal from the preamplifier is directly proportional to the number of ions  in the trap at the beginning of $t_\mathrm{Extract}$.	  
Additionally, the MOT can be turned on or off via an electronic shutter on one of the 589 nm retro-reflected beams during an adjustable percentage of the $t_\mathrm{Trap}$ and/or loading time $\mathrm{t_{Load}}$.

	The number of trapped ions was determined by calibrating the CEM to the observed photoionization from the MOT. The photoionization rate was  measured using the methods described in Ref.~\cite{Wipple:2001}. This calibration is accurate to an order of magnitude.

	Operating the LPT in the presence of the magnetic field produced by the anti-Helmholtz coils used to generate the MOT was experimentally established to have little impact on the trapping and detection of ions. Similarly the LPT rf fields did not affect the number of cold MOT atoms in any measurable way.


\begin{figure}[t]
   \centering
\scalebox{1.1}{\includegraphics[width=3.3in]{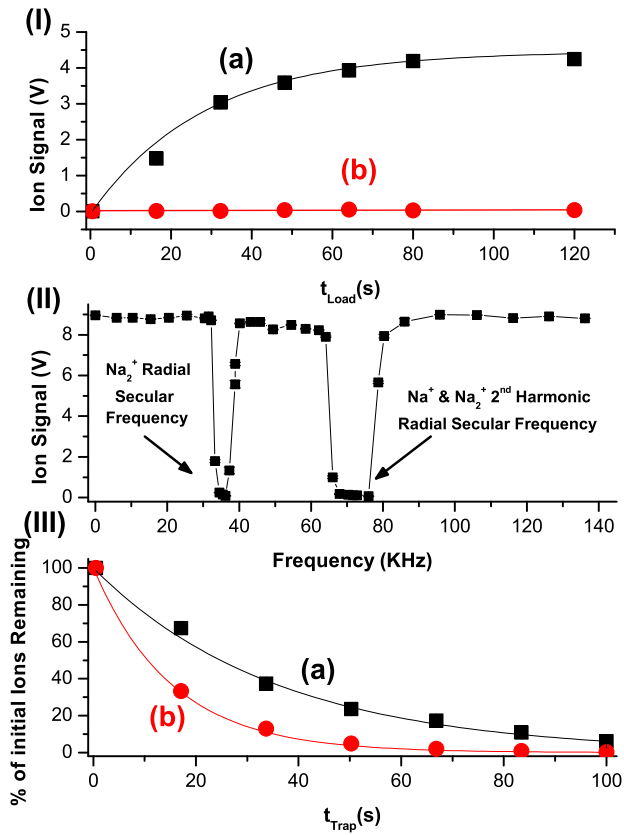} }
   \caption{(Color online) (I) (a)  \ce{Na2+} and the photodissociated \ce{Na+} produced by the MOT. (b)The null ion signal after mass selective resonant quenching applied at the \ce{Na2+} radial secular frequency. (II) A radial ac quenching field frequency scan of \ce{Na2+} and \ce{Na+} ions produced by the MOT due to AI. We find resonances at the \ce{Na2+} first harmonic at $36 \pm 1$ kHz, the \ce{Na2+} second harmonic at $71 \pm 1$ kHz, and the first harmonic of \ce{Na+} at $75 \pm 1$ kHz. (III) An illustration of the ac side-effect heating without the MOT. (a) Relative trap loss of \ce{Na+} ions without the ac quenching field applied to the trap segments of the LPT and (b) with the ac field set at the radial secular frequency of \ce{Na2+} ions at an amplitude $V_\mathrm{sec} = 1.8$ V. (The error bars are smaller than the data points).}
   \label{fig:Quench}
\end{figure}

\subsection{Trap loss mechanisms}
\label{sec:Trap loss mechanism}

The LPT can only trap ions whose energy is below the trap depth.The radial and the axial trap depths  \cite{Wineland:92} are defined as
 \begin{equation}
D_\mathrm{Radial} = \frac{e q_1 V_{\mathrm{rf}}}{4}-\frac{e \eta V_\mathrm{end} r_0^2}{2 z_0^2}
\end{equation}
and
\begin{equation}
D_\mathrm{Axial} =\eta eV_{\mathrm{end}}
\end{equation}
The LPT stability parameter is defined as $q_1 = \frac{4eV_{\mathrm{rf}}}{m_Ir_0^2\Omega^2}$ \cite{Goodman:12}.

For most of the results presented in Sec.~\ref{sec:Results}, when $q_1 \approx 0.32$, $D_\mathrm{Radial} \approx 2$ eV. Similarly, during the trapping time $D_\mathrm{Axial}$ was always $\approx 5$ eV. When an ion's mean secular energy rises above the trap depth due to heating within the LPT it will evaporate from the LPT resulting in trap loss. 

Heating mechanisms, which can result in ions being lost from the trap,  have to be overcome by the sympathetic cooling to achieve low ion temperatures. Trapped ions decay exponentially from the LPT, and ions which have been sympathetically cooled have been demonstrated experimentally to have longer lifetimes due to elastic scattering and non-radiative charge-exchange collisions \cite{Cote:2000} with cold neutral atoms \cite{Ravi:2012, Green:2007, Major:1968}.

	 Ion collisions cause energy to be  exchanged  between their micromotion and their secular motion, a process called atom-ion (if the collision is between the ion(s) and atoms) or ion-ion (between co-trapped ions) rf heating \cite{Schwarz:2008,Blumel:1989, Ryjkov:2005}. Another ion trap heating mechanisms is excess micromotion heating \cite{ZipkesPRL:2010,Berkeland:1998}. It is caused by imperfections  in the construction or the alignment of the trap electrodes, the relative phase of the electric fields, and stray fields present within the trapping region due to other electric devices as well as charge build up on insulating materials in the chamber. In addition to testing trap lifetime extension via sympathetic cooling, we experimentally found that increasing individual heating mechanisms reduced trap lifetimes in the LPT.

 Since ion-ion and atom-ion rf heating are an inevitable by-product of ion trapping in an LPT and are dependent on the value of the $q_1$ stability parameter \cite{Schuessler:05}, we had to experimentally determine the most favorable  $V_\mathrm{rf}$ for our LPT. In the absence of any cooling, we scanned $V_\mathrm{rf}$ and found the largest ion signal and longest trap lifetime corresponded to $q_1=0.5$. Similar results were found in Ref. \cite{Hasegawa:06}.  This is likely due to the balance of competing factors such as single ion stability, atom-ion rf heating, ion-ion rf heating, and trap depth. We found that with a $q<0.5$ reduced heating (as was seen in simulations performed in Ref.~\cite{Goodman:12}), but $q<0.2$ would significantly reduce the trap lifetime, likely due to the reduced trap depth.  Therefore, in trap lifetime measurements, which comprise the majority of the results in Sec.~\ref{sec:Results}, we use $q_1 \approx 0.32$.

	 In the presence of 589 nm light, \ce{Na} MOT atoms (uniquely among the alkali metal atoms but commonly in alkaline earth atoms) are additionally subject to photoassociative ionization (AI) reactions which produce \ce{Na2+} molecular ions  \cite{Gould:1988,Julienne:1991}. \ce{Na+} ions are produced from these \ce{Na2+} ions via two mechanisms: resonant photodissociation caused by the 589 nm photons and collisional photodissociation caused by collisions with excited Na(3p) atoms \cite{Tapalian:1994}.  As a result  the \ce{Na} MOT is itself a source of \ce{Na2+} and \ce{Na+} ions. Since the time of flight to the  CEM is not resolved for the two ionic species, any charged particles (atomic or molecular) introduced during $t_\mathrm{Trap}$  increases the ion signal. These  MOT born ions due to AI can interfere with the sympathetic cooling experiment by thermalizing with the  ion sample under study through Coulomb interactions, ultimately resulting in increased ion-ion rf heating.

	To quench unwanted ions from the LPT, mass selective resonant excitation was implemented \cite{Drakoudis:2006,Fortson:69}.  Figure ~\ref{fig:Quench}(I)(a) shows the \ce{Na2+} and the photodissociated \ce{Na+} produced during $t_\mathrm{Trap}$ from the MOT. An external ac field was introduced on the rf trap segments in a quadrupole configuration set at the radial secular frequency  of the trapped \ce{Na2+} ions. The amplitude of the external ac quenching field ($V_{\mathrm{sec}}$) is set  much lower than that of the rf driving field amplitude $V_\mathrm{rf}~ (V_{\mathrm{sec}}\ll 5 \%$  of  $V_\mathrm{rf})$ to prevent any disturbance of the trapping potential. The frequency of the external ac quenching field was scanned while monitoring the ion signal with the CEM  as shown in Fig.~\ref{fig:Quench}(II). Ions are ejected from the trap when their secular frequency resonates with the applied external ac field because the ions' energy is resonantly driven above the trap depth. Our experimental secular frequency measurements showed good agreement with \textsc{simion} simulations. For example, the values for \ce{Na+} $\omega_\mathrm{Axial}$ and $\omega_\mathrm{Radial}$ found via simulations were $\approx 36$ kHz and $\approx 76$ kHz; experimentally we found them to be $35 \pm 1$ kHz and $75 \pm 1$ kHz respectively.

	Although this technique sufficiently quenched extraneous ions, it presented a new heating mechanism referred to as ac side-effect heating \cite{Goodman:12}. When mass selective quenching is implemented resonantly at the \ce{Na2+} radial secular frequency, the motion of the trapped \ce{Na+} ions is perturbed.  As shown in Fig.~\ref{fig:Quench} (III), when the MOT was turned off during $t_\mathrm{Trap}$, the presence of the ac quenching field on the trap segments leads to additional trap loss. 
 
The ions created from AI within the MOT prevents us from studying a small number of ions. Even when quenching is on, the CEM measures a background signal from the AI ions. In order to keep a high signal to background ratio, the minimum initial number of \ce{Na+} ions loaded in the trap ranged from $10^3 - 10^4$.

The amplitude of the secular field, $V_{\mathrm{sec}}$, must be chosen to balance, sufficiently quenching the \ce{Na2+} ions and minimizing ac side effect heating. At the value of $q_1$ for \ce{Na+} that we used ($q_1 \approx 0.32$) we found a minimum value of $V_{\mathrm{sec}}\approx 1~$V was required to sufficiently quench the \ce{Na2+} background signal. We found that when $V_{\mathrm{sec}}$ was increased from this minimum value by as little as 20$\%$, the ac side effect heating would begin to overwhelm the sympathetic cooling from the MOT.

Both $q_1$ and $V_{\mathrm{sec}}$ are inversely dependent on the mass of the species trapped. Since \ce{Na2+} is twice the mass of \ce{Na+}, it experiences half the $q_1$ and half the trap depth experienced by \ce{Na+}. Therefore, as  $q_1$ of \ce{Na+} is lowered, the minimum necessary $V_{\mathrm{sec}}$ is also lowered; as a result the ac side-effect heating is also lowered. This gives further motivation for using $q_1<0.5$. Experimentally, we found that if $q_1<0.2$ the \ce{Na2+} trap depth and stability parameter are so low that the minimum necessary $V_{\mathrm{sec}}\approx 0~$V, i.e \ce{Na2+} is not trapped.


\section{RESULTS}
\label{sec:Results}
\begin{figure}[t]
   \centering
\scalebox{1.1}{\includegraphics[width=3.3in]{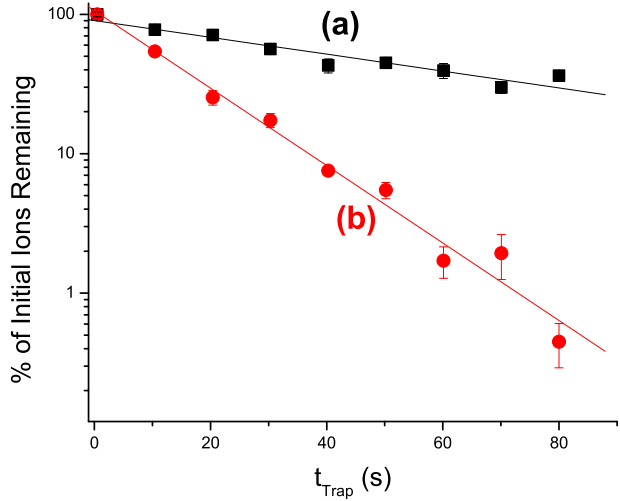} }
   \caption{(Color online). \ce{Na+} decay curves on a semilog scale have an exponential decay showing the difference in trap loss between ions that were sympathetically cooled by the MOT (a) and ions that were not exposed to the MOT (b). (The error bars are smaller than the data points where they are not visible).}
   \label{fig:Cooling1}
\end{figure}

\begin{figure}[t]
   \centering
\scalebox{1.1}{\includegraphics[width=3.3in]{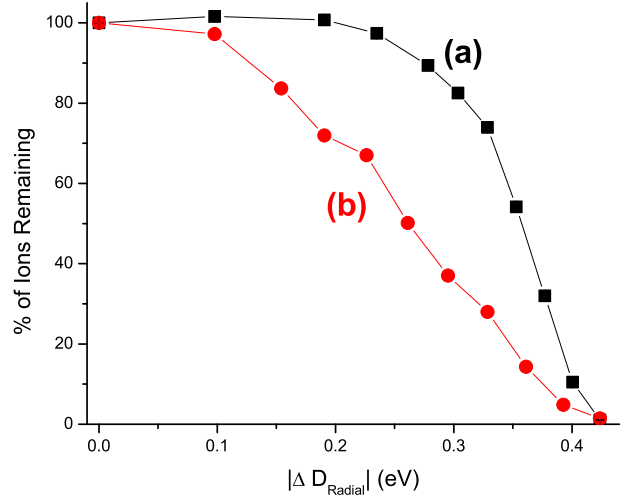} }
   \caption{(Color online). Fraction of ions remaining in the trap, at an initial $D_\mathrm{Radial} \approx 0.6$ eV, as a function of a sudden change in trap depth ($\Delta D_\mathrm{Radial}$) before extraction after a fixed $t_\mathrm{Trap}$ = 5 s. Curve (a) is with MOT cooling. Curve (b) is without MOT cooling. (The error bars are smaller than the data points). }
   \label{fig:qdrop}
\end{figure}

\begin{figure}[t]
   \centering
\scalebox{1.0}{\includegraphics[width=3.3in]{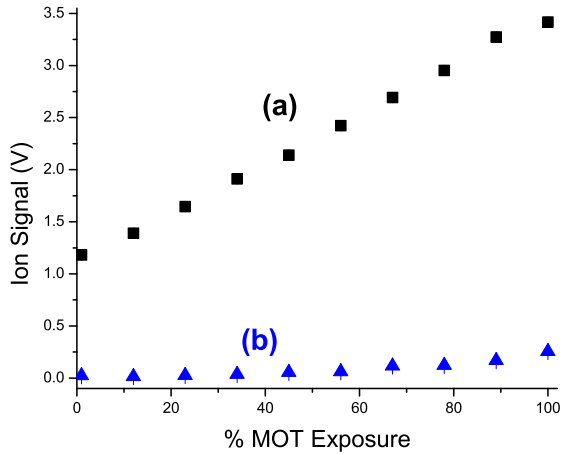} }
   \caption{(Color online). Ion signal (proportional to the number of ions) plotted against the fraction of  a fixed $t_\mathrm{Trap}$ = 8 s during which the ions were exposed to the MOT. In curve (a) the ions signal is seen to increase linearly as MOT exposure approaches 100$\%$  of the fixed trapping time. In curve (b) the background reading  shows effective quenching of  AI MOT produced ions. (The error bars are smaller than the data points where they are not visible). }
   \label{fig:percent_mot}
\end{figure}

\begin{figure}[t]
   \centering
\scalebox{1.1}{\includegraphics[width=3.3in]{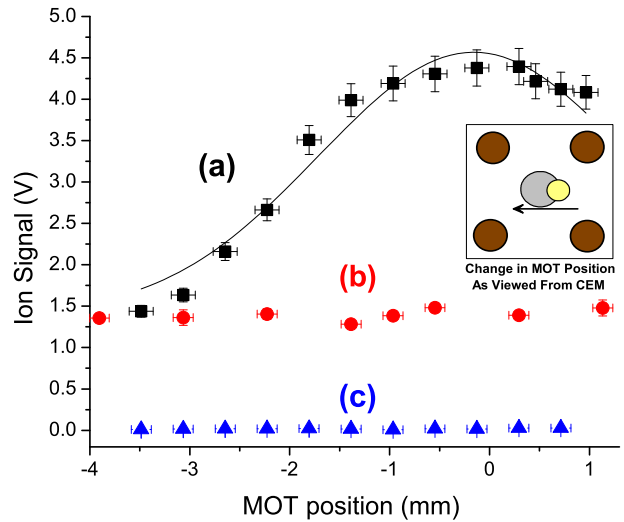} }
   \caption{(Color online). (a) Ion signal (proportional to the number of ions) is plotted against the translated position of the MOT, which is a caused by an applied shim coil current. Position $=0$ mm corresponds to the point of maximum ion signal. A Gaussian fit of curve (a)  gives a full width at half maximum of $3.2 \pm 0.2$ mm. (b) Effect of the shim coil on the trapped ions when the MOT is not present. (c) Background MOT born AI ion signal showing that translating the MOT across the LPT axis does not affect the quenching process. Inset: Shows the MOT moving  (+) to (-) as (from right to left) as viewed from the CEM. (The error bars are smaller than the data points where they are not visible).}
   \label{fig:Overlap}
\end{figure}

	Since ground state \ce{Na+} does not have an optically accessible transition, direct fluorescence based Doppler temperature measurements were not possible.  Instead,  four different indirect measurements were taken to demonstrate sympathetic cooling of \ce{Na+} ions by the \ce{Na} MOT.

	Figure~\ref{fig:Cooling1} shows a semilog plot of a typical ion lifetime measurement with and without sympathetic cooling by the MOT. The ions were initialized from the background \ce{Na} gas resulting in an initial time averaged mean ion cloud kinetic energy of $\approx 1 \mathrm~{eV}$ (according to \textsc{simion} simulations).  The external quenching of \ce{Na2+} ions was implemented during this process and any residual background signal was subtracted from the final experimental values to eliminate any contributions from ions produced during $t_\mathrm{Trap}$. As shown in Fig.~\ref{fig:Cooling1}, \ce{Na+} ions that were cooled by the \ce{Na} MOT stayed in the trap longer. Similar results were obtained for a smaller number of ions in Ref.~\cite{Ravi:2012} with \ce{Rb+} ions and a \ce{Rb} MOT.

The shape of the decay curves shown in Fig.~\ref{fig:Cooling1} are typical whether the ions are loaded via REMPI, from the MOT or the background gas. Therefore the temperature of the source of the neutrals from which the ions are produced has little effect on the trap lifetime or the final temperature, as predicted in Ref.~\cite{Goodman:12}.


The second test used to demonstrate sympathetic cooling measures the trap loss as a function of changing trap depth. When the ion cloud is cooled by the MOT the energy distribution of the ion cloud changes. Therefore, a hotter ion cloud should yield a larger fraction of ions lost after a sudden drop of the LPT's trap depth \cite{Chen:2007}.

After  $t_{\mathrm{Trap}}$ = 5 s with a $D_\mathrm{Radial}$ of 0.6 eV, the radial trap depth was lowered suddenly by $\Delta D_\mathrm{Radial}$ for 10 ms duration ($t_{Drop}$)  by reducing $V_{rf}$ immediately prior to extraction. After suddenly lowering $V_{rf}$, the ions were detected using the CEM.  As $\Delta D_\mathrm{Radial}$ is increased, the ions that are not cooled  begin to evaporate from the trap at a much smaller $\Delta D_\mathrm{Radial}$ than when the ions are sympathetically cooled  (Fig.~\ref{fig:qdrop}).

This experiment (Fig.~\ref{fig:qdrop}) was conducted at an initial \ce{Na+} $q_1$ = 0.18 value, for which the AI-produced \ce{Na2+} and \ce{Na+} ions are not captured in the LPT and therefore secular quenching was not necessary. However, this test produced similar results at various $q_1$, $t_{Drop}$ and $t_{Trap}$ values.

	 The third method employed to show sympathetic cooling was by changing the percent of trapping time an ion sample was exposed to the MOT during a fixed $t_\mathrm{Trap} = $ 8 s. The ion signal increases linearly  as a function of  increasing MOT exposure time, i.e., increased exposure time leads to a larger fraction of ions cooled below the trap depth  (Fig.~\ref{fig:percent_mot}). 

	 The overlap of the MOT with the ion cloud, which was demonstrated to have a significant effect on sympathetic cooling in Refs  \cite{Goodman:12, Schmid:2010}, was tested and provided a fourth and final test to demonstrate cooling. Moving the MOT with respect to the ion cloud was accomplished by using a magnetic shim coil. The trapped ion signal is obtained after a fixed $t_\mathrm{Trap} =$ 7 s. As portrayed in Fig.~\ref{fig:Overlap}(a), the ion signal reached a maximum as the MOT was translated across the $x_1-x_2$ plane. Although the ion cloud cannot be optically imaged, the relative MOT position at which the maximum ion signal occurs is likely where the MOT is concentric with the ion cloud. A Gaussian fit of curve (a) Fig.~\ref{fig:Overlap} yields a full width at half maximum of $3.2 \pm 0.2$ mm. This measurement can be interpreted as an upper bound of the size of the ion cloud in the MOT translation direction.  The shim coil itself does not dramatically affect the ion signal [see Fig.~\ref{fig:Overlap} (b)] or the  resonant quenching of extraneous ions [see Fig.~\ref{fig:Overlap} (c)].

\section{CONCLUSION}
\label{sec:Conclusions}

	 We  demonstrated sympathetic cooling of  \ce{Na+} ions by a cold Na MOT in a hybrid trap. Since the Na MOT also produces \ce{Na2+} and \ce{Na+} ions via photoassociative ionization and subsequent photodissociation, measures were taken to quench this external ion production in order to demonstrate a clear cooling effect, which we observed despite this experimental obstacle.

	 Evidence of sympathetic cooling by cold MOT atoms was investigated using four different methods.  Difference in trap lifetime,
	 trap loss due to changing trap depth, variable MOT exposure time and MOT overlap were tested experimentally and the results support sympathetic cooling of  \ce{Na+} ions by an equal mass Na MOT as hypothesized. Because of our previously published work in simulating this system \cite{Goodman:12}, this result was surprising to us (but not inconsistent) that the hybrid trap is able to effectively cool a relatively large number of co-trapped ions. 

\section{Acknowledgments}
\label{sec:Acknowledgments}

	 We would like to acknowledge support from the NSF under Grant No. PHY0855570. One of us (F.A.N.)  would like to thank the University of Connecticut group for their hospitality during numerous visits. We also thank Jian Lin and Oleg Makarov for their preliminary work on the ring-dye laser and the Na MOT.


\bibliography{References}
\end{document}